\documentclass{amsproc}

\newtheorem{theorem}{Theorem}[section]
\newtheorem{lemma}[theorem]{Lemma}

\theoremstyle{definition}
\newtheorem{definition}[theorem]{Definition}
\newtheorem{example}[theorem]{Example}
\newtheorem{xca}[theorem]{Exercise}

\theoremstyle{remark}
\newtheorem{remark}[theorem]{Remark}

\numberwithin{equation}{section}

\newcommand{\abs}[1]{\lvert#1\rvert}

\newcommand{\blankbox}[2]{%
  \parbox{\columnwidth}{\centering
    \setlength{\fboxsep}{0pt}%
    \fbox{\raisebox{0pt}[#2]{\hspace{#1}}}%
  }%
}

\begin{document}

\title{ Entanglement beyond subsystems }

\author{ L. Viola }
\address{Los Alamos National Laboratory, Los Alamos, New Mexico 
87545 }
\email{lviola@lanl.gov}
\thanks{Research sponsored by the US DOE. L.V. gratefully acknowledges 
support from the Los Alamos Office of the Director through a J.R. 
Oppenheimer Fellowship.  This paper is based on a talk delivered at the 
{\em Coding Theory and Quantum Computing Conference}, 
University of Virginia, Charlottesville, May 20-24, 2003. }

\author{ H. Barnum}
\address{Los Alamos National Laboratory, Los Alamos, New Mexico 87545 }
\email{barnum@lanl.gov}

\author{ E. Knill}
\address{National Institute of Standards and Technology, Boulder, 
Colorado 80305 }
\email{knill@boulder.nist.gov}
 
\author{ G. Ortiz}
\address{Los Alamos National Laboratory, Los Alamos, New Mexico 87545 }
\email{ortiz@viking.lanl.gov}

\author{ R. Somma }
\address{Los Alamos National Laboratory, Los Alamos, NM 87545, USA}
\email{somma@viking.lanl.gov}

\date{January ???, 2004 and, in revised form, ???, 2004.}



\begin{abstract}
We present a notion of generalized entanglement which goes beyond the 
conventional definition based on quantum subsystems.  This is accomplished
by directly defining entanglement as a property of quantum states 
relative to a distinguished set of observables singled out by Physics.
While recovering standard entanglement as a special case, our notion
allows for substantially broader generality and flexibility, being
applicable, in particular, to situations where existing tools are not
directly useful.
\end{abstract}

\maketitle

.

\section{Introduction} 

Since the early days of quantum mechanics, the 
concept of {\it entanglement} has proved a continuous source of 
physical, mathematical, and philosophical challenges.  Interest in 
entanglement has heightened dramatically in recent years, the 
characterization and quantification of entanglement being nowadays 
one of the most active research areas within the emerging science 
of Quantum Information (QI)~\cite{NC}.  
So, what is the point about entanglement?  Quoting from a landmark 
1935 paper by Schr\"odinger, where the term entanglement was first 
introduced~\cite{S} (see also~\cite{EPR}), 
\begin{quotation}
``...This is the point. 
Whenever one has a complete expectation catalog - a maximum total
knowledge - a $\psi$ function - for two completely separated 
bodies,... then one obviously has it also for the two bodies together...
But the converse is not true. Maximal knowledge of a total system 
does not necessarily include total knowledge of all its parts, not
even when these are fully separated from each other and at the
moment are not influencing each other at all.'' 
\end{quotation}

This can be interpreted as pointing out the possibility that some pure
states of composite quantum systems -- precisely those Schr\"odinger
calls ``entangled'' -- have mixed {\it reduced} states on their
constituent subsystems.  It is exactly this aspect of entanglement
that we propose to generalize beyond the subsystem framework. Of
course, a lot more has been learned by now about the properties and
significance of such states from both a physical and an
information-theoretic perspective.  In particular, because no pure
state of a classical system can be correlated, the correlations
present in a pure entangled state are intrinsically non-classical, as
witnessed by the fact that the presence of entanglement is a necessary
condition for the violation of Bell inequalities~\cite{BE,PS}.  In
addition, entangled states are known to provide the defining resource
for quantum communication, enabling non-classical protocols such
quantum teleportation and leading to provable improvements in the
efficiency of various multipartite tasks~\cite{BU}.  Finally, although
a full assessment of the role of entanglement in quantum computational
speed-up remains elusive to date, entangled states involving
unboundedly many qubits turn out to be necessary for efficient
pure-state quantum algorithms~\cite{JL}.

In spite of such progress, there have been a number of signs recently
that the theory of entanglement developed so far is too restrictive 
to be applicable to the full variety of both QI and physical settings 
one might like to consider.  In particular, the conventional approach
to entanglement focuses on analyzing how a quantum system is made up
from constituent subsystems, and implicitly incorporates the assumption
that the latter are operationally {\it distinguishable}.  Compelling 
motivations for critically reconsidering the resulting entanglement 
notion come from situations where the operational access to the system 
of interest is inherently restricted by physical constraints.  A first 
example is offered by condensed-matter systems, where the consequences
of non-trivial (bosonic or fermionic) {\it quantum statistics} must be 
taken into account, making the direct identification of particles as
``entanglable'' subsystems problematic.  A second example may occur in 
systems which are subject to {\it superselection rules}, which restrict
the allowed transformations on the system's state space and effectively 
cause sectors of the latter to be operationally unreachable.  Both 
situations are being actively investigated in the QI literature, and 
no attempt will be made here to provide a complete account.  While
different solutions have been examined thus far, they essentially all 
rely on either appropriately redefining the relevant subsystems to be
used as primitive blocks, or the applicable measure for quantifying 
the accessible correlations -- leaving, however, the underlying 
entanglement concept unchanged.  We refer the reader 
to~\cite{ON,EK,ZF} and~\cite{BW,VC} for representative discussions 
of entanglement for indistinguishable particles and in the presence 
of superselection rules, respectively. 
 
Our approach takes a different route, bypassing the need of a subsystem 
decomposition in the first place, and reformulating entanglement
as a property of quantum states which directly depends on physical
{\it observables}.  The investigation of the resulting notion of 
{\it generalized entanglement} (GE) was undertaken in~\cite{BKOV} and 
continued in~\cite{BKOSV}.  In the following, we further highlight 
some properties and implications of such a perspective.

\section{Entanglement and preferred subsystems}

Let $S$ be a quantum system with associate state space ${\mathcal H}$,
dim (${\mathcal H})=d$, for some $d \in {\mathbb N}$.  For simplicity, 
we shall focus on considering {\it pure} states of $S$ throughout.  
A subsystem of $S$ can be defined in general as a {\it factor} (in the 
tensor product sense) of a subspace ${\mathcal C} \subseteq 
{\mathcal H}$~\cite{KLV}.  In the simplest case where ${\mathcal C}=
{\mathcal H}$, a subsystem decomposition of $S$ is described by a
tensor product structure on ${\mathcal H}$,
\begin{equation}
{\mathcal H} \simeq \bigotimes_\ell {\mathcal H_\ell},
\label{tps}
\end{equation}
where ${\mathcal H_\ell}$ is the state space of the $\ell$th subsystem.
Whenever a multipartition of the form (\ref{tps}) is defined over a {\it 
proper} subspace of  ${\mathcal H}$, an {\it encoded} tensor product 
structure is obtained, and the underlying subsystems are sometimes 
referred to as ``logical'' or ``virtual'' subsystems~\cite{Z}. 

Even in the simplest setting described by Eq. (\ref{tps}), it is essential 
to realize that, {\it a priori}, the number of inequivalent subsystem 
partitions that ${\mathcal H}$ can support may be large, reflecting the 
fact that many possibilities of expressing $d$ as a product of integers 
arise in general \footnote{See \cite{Z} for a formal definition of 
equivalent factorizations and the resulting notion of 
tensor-product-structure manifold.}.
Accordingly, entanglement of a state in ${\mathcal H}$ is unambiguously 
defined only once a preferred subsystem decomposition is selected -- 
which physically corresponds to a specific way of viewing $S$ as made 
up from constituent parts.  Relative to a distinguished multipartite
structure, a pure state $|\psi\rangle \in {\mathcal H}$ is entangled iff 
is {\it not}  expressible as a product of states in $ {\mathcal H_\ell}$. 
Given $|\psi\rangle$, the state of the $\ell$th subsystem is determined 
by the reduced density operator $\rho_\ell$ resulting from the partial 
trace operation over the remaining subsystems.  Thus, an equivalent 
characterization of standard pure-state entanglement is that $|\psi\rangle$ 
is entangled iff $\rho_\ell$ is {\it mixed} for some $\ell$: entangled 
pure states are exactly those which appear mixed to observers whose 
operational access to $S$ is restricted to {\it local observables} 
acting on only a subsystem at a time. 

The above observations suggest that the standard definition of entanglement 
can be equivalently phrased in terms of the observables used to measure the 
system and describe its states, and that perhaps this could serve as a starting 
point for an operational characterization of entanglement under more general 
circumstances.  A natural path for capturing such an intuition is to 
relate the emergence of a preferred subsystem decomposition to be used to 
the set of available interactions and measurement capabilities. Such a path, 
which naturally extends the observale-based definition of subsystems in 
terms of associative algebras introduced in~\cite{KLV,VKL,Z}, has 
been recently investigated by Zanardi {\it et al.}~\cite{ZLL}.  The essence 
of our GE approach, as we shall see, is instead to redefine entanglement so 
as to make it {\it directly dependent upon a physically relevant set of 
observables} -- irrespective of and without reference to a preferred way
for partitioning $S$ into subsystems.

\section{The notion of generalized entanglement}

The key realization underlying the GE approach is that the distinctive 
features of entanglement are determined by the expectation values of a
distinguished {\it subspace of observables} of $S$. The latter may directly
incorporate an operational restriction, such as limited means for 
controlling and measuring the system, or more generally reflect an insight
or condition dictated by Physics.  Let such a preferred observable 
subspace be denoted by $\Omega$.    The steps leading to GE may be 
summarized as follows:

\begin{itemize}
\item Pure states are unentangled (in the conventional sense) iff every
subsystem state is pure;
\item States can be represented as positive linear functionals on operator
spaces, induced by a density operator according to the trace map.  Formally, 
given a state of $S$ described by $\rho$, this completely specifies a positive 
linear functional $\lambda$ on the space End(${\mathcal H}$) of all operators 
on ${\mathcal H}$ via the action $$ \lambda (X) = \text{tr} (\rho X), 
\hspace{5mm} X \in \text{End}({\mathcal H}), $$
such that $\lambda(X^\dagger X) \geq 0$ and $\lambda({\mathbb I}) =1$. 
Such a state $\lambda$ can be restricted to $\Omega \subseteq \text{End}
({\mathcal H})$, giving a reduced state $\omega = \lambda | \Omega$ which 
only determines the expectations of operators in $\Omega$.  
If, additionally, $\Omega $ is closed under Hermitian conjugation, 
we can define a reduced density operator $\rho_{red} \in 
\Omega$ associated with the state by the relationship
$$ \omega (Y) = \text{tr} (\rho Y) = \text{tr} (\rho_{red} Y), 
\hspace{5mm} Y \in \Omega ; $$ 
\item The set ${\mathcal R}$ of $\Omega$-reduced states is convex, 
that is, closed under probabilistic combinations;
\item A state in ${\mathcal R}$ is pure iff it is extremal, that is, it
cannot be expressed as a non-trivial convex combination of elements in 
${\mathcal R}$.
\end{itemize}

We are thus led to the following: 

\begin{definition}
A pure state $\rho=|\psi\rangle\langle \psi|$ of $S$ is {\it generalized 
unentangled relative to the distinguished set of observables} $\Omega$ if 
its reduced state is pure, and generalized entangled 
otherwise\footnote{Similarly, a mixed state $\rho$ of $S$ is generalized 
unentangled relative to $\Omega$ if it can be written as a proper convex 
combination of generalized unentangled pure states.}.
\end{definition}

While the above definition is applicable to an arbitrary linearly closed 
operator set $\Omega$, a relevant situation which is often encountered 
in physical applications occurs when $\Omega$ is a {\it Lie algebra} of 
operators acting on ${\mathcal H}$. 
 
\subsection{Lie-algebraic framework and Lie-algebraic purity}

Let us focus on the case where $\Omega = {\mathfrak{h}}$ is a semisimple 
Lie algebra of operators acting {\it irreducibly} on the representation
space ${\mathcal H}$.  In order to make the connection with physical 
observables more transparent, we also assume that ${\mathfrak{h}}$ is a
real Lie algebra consisting of Hermitian operators, with a modified 
Lie-bracket operation defined as follows:
$$ [X,Y] = i (XY - YX), \hspace{5mm} X,Y \in  {\mathfrak{h}}.$$
Accordingly, the Lie group $G$ generated by ${\mathfrak{h}}$ will be 
obtained via the modified exponential map $X \mapsto e^{iX}$. The reader 
is referred to~\cite{G,HA,HU} for relevant background on  Lie representation 
theory.  The general case of a reducible action of ${\mathfrak{h}}$ on 
${\mathcal H}$ is addressed elsewhere (see for instance ~\cite{SOBKV}).

Under the above assumptions, one of the key results established 
in~\cite{BKOV} (see Theorem 14, part (3)) is the identification of the 
generalized unentangled pure states as the {\it generalized coherent states}
(GCSs) associated with the Lie algebra ${\mathfrak{h}}$ (see also 
\cite{K} for a related characterization motivated by geometric invariant 
theory).  A review of the theory underlying GCSs and their main physical 
applications may be found in~\cite{PR,ZFG}.  Recall that GCSs are obtained 
by extending the definition of canonical (harmonic-oscillator) coherent 
states in terms of a displaced reference state within a general Lie-algebraic 
framework.  Beside the Lie algebra ${\mathfrak{h}}$ with associated dynamical 
group ${G}$ acting on ${\mathcal H}$, the definition of GCSs requires the 
specification of a normalized reference state 
$|\Psi_0\rangle \in {\mathcal H}$ which, following Gilmore's 
construction, is chosen as an extremal (lowest-weight) state of the 
representation.
Knowledge of ${G}$ and $|\Psi_0\rangle$ determines the subgroup 
${G_0} \subseteq G$ of elements that preserve the reference state up to a 
phase factor. The manifold of GCSs associated to $({\mathfrak{h}}, 
{\mathcal H}, |\Psi_0\rangle)$ is then defined by the {\it orbit of the 
reference state under coset elements} in ${G}/{G_0}$ that is,
\begin{equation}
\text{GCS}_{( {\mathfrak{h}}, {\mathcal H}, |\Psi_0\rangle)} = 
\{ {\mathcal D} |\Psi_0\rangle\, |\,{\mathcal D} \in {G}/{G_0} \}.
\label{gcs} 
\end{equation}
Accordingly, all generalized unentangled states are GCSs and are in the 
orbit of a (unique) minimum weight state of ${\mathfrak{h}}$ under the 
action of the Lie group. 

Given a pure state $|\psi\rangle \in {\mathcal H}$, one would like to
obtain a quantitative way for relating the properties of generalized 
unentanglement, generalized coherence, and purity with respect to the 
observable subspace.  A natural procedure is to consider the projection
of $|\psi\rangle\langle \psi|$ onto ${\mathfrak{h}}$, as the latter 
completely determines the expectation values of operators in the Lie
algebra.  This motivates the following:  
\begin{definition}
Let $\{x_i\}$ be a Hermitian and commonly-normalized orthogonal ($\text{tr} \
x_ix_j \propto \delta_{ij}$) basis for ${\mathfrak{h}}$. For any $|\psi\rangle 
\in {\mathcal H}$, the purity of $|\psi\rangle$ relative to ${\mathfrak{h}}$ 
(or ${\mathfrak{h}}$-purity) is
\begin{equation}
P_{ \mathfrak{h}}( |\psi\rangle )= \sum_i 
(\text{tr} |\psi\rangle\langle \psi| x_i)^2 =
\sum_i |\langle \psi|x_i | \psi\rangle|^2.
\label{ph}
\end{equation}
\end{definition}

\begin{remark}
Let ${\mathcal P}_{\mathfrak{h}}$ denote the projection map onto 
${\mathfrak{h}}$.  As defined, the ${\mathfrak{h}}$-purity of a state 
is the square length of the projection ${\mathcal P}_{\mathfrak{h}} 
(|\psi\rangle\langle \psi|)$ according to the trace-inner-product norm.
Note that, in principle, different operator norms could be used, 
resulting in different purity functionals for fixed ${\mathfrak{h}}$.
\end{remark}

\begin{remark}
Let $D \in G$ be an arbitrary group transformation.  Then 
${P}_{\mathfrak{h}}(|\psi\rangle )$$ = \tilde{ P}_{\mathfrak{h}}
(|\psi\rangle )$, where the latter is calculated by replacing the 
operators $x_i$ with $D^\dagger x_i D$ in Eq. (\ref{ph}). Because 
$\tilde{ P}_{\mathfrak{h}}(|\psi\rangle) 
= { P}_{\mathfrak{h}}( D |\psi\rangle) $, the ${\mathfrak{h}}$-purity 
is invariant under group transformations, as is desirable on physical 
grounds.
\end{remark}

With the above definitions, the main features of GE within the 
Lie-algebraic framework are summarized by the following:

\begin{theorem}
The following characterizations are equivalent for an irreducibly 
represented Lie algebra ${\mathfrak{h}}$ on ${\mathcal H}$: 
\begin{enumerate}
\item $\rho$ is generalized unentangled relative to ${\mathfrak{h}}$.
\item $\rho=|\psi\rangle\langle\psi|$ with $|\psi\rangle$ the unique 
ground state of some Hamiltonian $H$ in ${\mathfrak{h}}$.
\item $\rho=|\psi\rangle\langle\psi|$ with $|\psi\rangle$ a 
lowest-weight vector of ${\mathfrak{h}}$.
\item $\rho$ has maximum ${\mathfrak{h}}$-purity.
\end{enumerate}
\end{theorem}

A proof of this Theorem is available in~\cite{BKOV}.  We also refer the 
reader to the same paper for an extended discussion of generalizations of 
various information-theoretic notions (such as local maps and complexity
measures) to the Lie-algebraic setting and beyond.  In the following, we 
focus on illustrating some aspects of the relationship between 
conventional and generalized entanglement.

\section{Conventional entanglement revisited}

Given the quantum system $S$ and a pure state $|\psi\rangle \in {\mathcal H}$, 
the purity relative to the (real) Lie algebra of all traceless observables 
${\mathfrak{h}}={\mathfrak{su}(d)}$ spanned by an orthogonal, commonly 
normalized Hermitian basis $\{ x_1 \cdots x_{L} \}$, $L=d^2-1$, is, 
according to Eq. (\ref{ph}), given by:
\begin{equation}
\label{puritysubsystem}
P_{ \mathfrak{h}}( |\psi\rangle )= {\sf K} \sum_{\alpha=1}^{L} 
\langle x_{\alpha} \rangle^2 ,
\end{equation}
where the overall normalization ${\sf K}$ depends on the dimension $d$ and 
is determined by requiring that the maximum purity value is 1.  
For an orthonormal operator basis, 
{\sf Tr}$(x_\alpha x_\beta)=\delta_{\alpha, \beta}$, 
${\sf K}=d/(d-1)$, whereas in the case {\sf Tr}$(x_\alpha x_\beta)= d
\delta_{\alpha, \beta}$ (as for ordinary un-normalized Pauli matrices), 
${\sf K}=1/(d-1)$.  Because any quantum state $|{\psi}\rangle \in {\mathcal H}$ 
can be obtained by applying a group operator ${\mathcal D}$ to a reference 
state $|{{\sf ref}}\rangle$ (a lowest-weight state of ${\mathfrak{su}(d)}$), 
\begin{equation}
|{\psi} \rangle = {\mathcal D} |{{\sf ref}}\rangle , \hspace{5mm} 
{\mathcal D}=e^{i \sum_{\alpha} t_{\alpha} x_{\alpha}}, \;
t_{\alpha} \in {\mathbb C},
\end{equation}
any quantum state $|{\psi}\rangle$ is a GCS of ${\mathfrak{su}(d)}$, 
thus generalized unentangled relative to the {\em full} observable algebra:
$P_{ \mathfrak{h}}( |\psi\rangle )=1$ for all $|\psi \rangle $.

Let now assume that $S$ is composed of $n$ distinguishable susbsytems, 
corresponding to a state-space factorization of the form (\ref{tps}), 
that is, ${\mathcal H}\simeq \otimes_\ell {\mathcal H}_\ell$, 
with dim(${\mathcal H}_\ell)=d_\ell$, $\prod_\ell d_\ell =d$.  
In the conventional setting, distinguishability of the subsystems motivates 
the assumption of {\em local accessibility} to their individual state spaces. 
Thus, the set of {\em local observables}, consisting of operators which act 
on each subsystem independently, is physically distinguished. Let 
\begin{equation}
\mathfrak{h}_{loc}= \bigoplus_{\ell=1}^n \mathfrak{su}(d_\ell)
\end{equation}
denote the corresponding (real) Lie algebra of traceless local observables, 
acting irreducibly on ${\mathcal H}$.  Because standard unentangled pure 
states are exactly those for which all reduced states remain extremal, the GE 
notion relative to the local observable subspace coincides with the standard 
notion of pure state multipartite entanglement~\cite{BKOV}.  This may be 
explicitly appreciated by studying the local purity $P_{\mathfrak{h}}$ and 
its relation to the conventional subsystem purities determined by the 
reduced subsystem states. 

\subsection{Local purities vs subsystem purities}
An orthonormal basis which is suitable for calculating $P_{\mathfrak{h}}=
P_{\mathfrak{h}_{loc}}$ 
may be obtained by considering a collection of orthonormal bases 
$\{  x^\ell_{\alpha_1} \cdots x^\ell_{\alpha_{L_\ell}} \}$, $L_\ell=d_\ell^2-1$, 
each acting on the $\ell$th factor alone, that is, 
\begin{equation}
\label{Basis}
x^\ell_{\alpha_\ell} = \overbrace{ {\mathbb I}^1 \otimes {\mathbb I}^2 \otimes 
\cdots \otimes \underbrace{x_{\alpha_\ell}}_{\ell^{th}\ \mbox{factor}} \otimes 
\cdots \otimes {\mathbb I}^n}^{n\ \mbox{factors}} ,
\end{equation}
where ${\mathbb I}^\ell ={\mathbb I}/\sqrt{d_\ell}$.  Let also 
${\mathfrak{h}_\ell}=\text{span}\{ x_{\alpha_\ell}\}$ be the Lie algebra 
of traceless Hermitian operators acting on ${\mathcal H}_\ell$ alone, and 
$\rho_\ell ={\sf Tr}_{\ell'\not = \ell} (\{ |\psi\rangle\langle \psi|\})$ 
the reduced density operator describing the $\ell$th subsystem's state.  
We can then prove:
\begin{theorem}
For every pure state $|\psi\rangle \in {\mathcal H}$ 
the following identities hold:
\begin{enumerate}
\item For each $\ell$, the purity relative to the algebra ${\mathfrak{h}_\ell}$ 
is proportional to the conventional subsystem purity:
\begin{equation}
P_{{\mathfrak{h}_\ell}} ( |\psi\rangle ) = 
\frac{d_\ell}{d_\ell -1} \left(  {\rm{tr}} \rho_\ell^2 - \frac{1}{d_\ell}
\right).
\label{hellreduced}
\end{equation}
\item The purity relative to the local algebra ${\mathfrak{h}}$ is 
proportional to the average subsystem purity: 
\begin{equation}
P_{{\mathfrak{h}_{loc}}} ( |\psi\rangle ) =  
\frac{1}{1 -\frac{1}{n} \sum_{\ell=1}^n \frac{1}{d_\ell}} 
\left[ \frac{1}{n} \sum_{\ell=1}^n \left(
{\rm{tr}} \rho_\ell^2 - \frac{1}{d_\ell} \right) \right].
\label{hreduced}
\end{equation}
\end{enumerate}
\end{theorem}
\begin{proof}
The state of the $\ell$th subsystem may be represented as  
$$ \rho_\ell = \frac{\mathbb I}{d_\ell}  + \sum_{\alpha_\ell} 
\langle x_{\alpha_\ell} \rangle x_{\alpha_\ell},  $$
where $\langle x_{\alpha_\ell} \rangle 
={\rm tr} (\rho_\ell  x_{\alpha_\ell}) = {\rm tr} (|\psi\rangle\langle
\psi| x^\ell_{\alpha_\ell } )$ and the last equality follows from the 
definition of reduced density operator.  Thus, 
$$ {\rm tr}\rho_\ell^2= \frac{1}{d_\ell} + \sum_{\alpha_\ell} 
\langle x_{\alpha_\ell} \rangle^2  x_{\alpha_\ell}. $$ 
By combining the above relation with the definition of the 
$P_{{\mathfrak{h}_\ell}}$-purity, 
$$P_{{\mathfrak{h}_\ell}} (|\psi\rangle) =
\frac{d_\ell}{d_\ell -1} \sum_{\alpha_\ell=1}^{L_\ell} 
\langle x_{\alpha_\ell} \rangle^2 x_{\alpha_\ell},$$
Eq. (\ref{hellreduced}) follows. 

To derive the second equality, start from the definition of the 
$P_{{\mathfrak{h}}}$-purity which, using Eqs. (\ref{ph}) and 
(\ref{Basis}), reads: 
$$ P_{{\mathfrak{h}} }(|\psi\rangle )= {\sf K}' \sum_{\ell=1}^n
\sum_{\alpha_\ell=1}^{L_\ell} \langle x^\ell_{\alpha_\ell} \rangle^2 =
{\sf K}' \sum_{\ell=1}^n \frac{d_\ell-1}{d_\ell} 
P_{ {\mathfrak{h}}_\ell }(|\psi\rangle).$$
Clearly, the maximum value of the above quantity will be attained when, 
and only when, each of the terms ${\rm tr} \rho_\ell^2=1 \leftrightarrow$
$P_ {{\mathfrak{h}}_\ell }=1$ for all $\ell$, corresponding to a pure 
product state of the form $|\psi\rangle= \otimes_\ell |\psi_\ell
\rangle$.  This allows to determine the ${\sf K}'$-normalization factor as
$${\sf K}'=
\frac{1}{ n - \sum_\ell \frac{1}{d_\ell}  } = 
\frac{1}{ n \Big( 1 - \frac{1}{n} \sum_\ell \frac{1}{d_\ell} \Big) },$$
leading to the expression given in Eq. (\ref{hreduced}).
\end{proof}

\begin{remark}
Note that $1/d_\ell \leq {\rm{tr}} \rho_\ell^2 \leq 1$ for every $\ell$, 
corresponding to a totally mixed and pure reduced state, respectively.  
Thus, the quantity in round parentheses in Eq. (\ref{hellreduced}) 
varies between $0$ and $1-1/d_\ell$, consistently implying a 
$P_{{\mathfrak{h}}_\ell}$-purity range between $0$ and $1$.
\end{remark}

\begin{remark}
In the case of co-dimensional subsystems, $d_\ell=d_0 \forall \ell$, 
Eq. (\ref{hreduced}) takes the simpler form
$$ P_{{\mathfrak{h}}} ( |\psi\rangle ) = 
\frac{d_0}{d_0 -1}\left[ \frac{1}{n} \sum_{\ell=1}^n 
\left(  {\rm{tr}} \rho_\ell^2 - \frac{1}{d_0}\right)\right]
= \frac{1}{d_0-1}\left( \frac{d_0}{n} \sum_{\ell=1}^n 
 {\rm{tr}} \rho_\ell^2 -1 \right) .$$
\end{remark}

According to the above Theorem, states corresponding to maximal local purity 
(i.e., generalized unentangled) are exactly the set of standard separable
states.  Note, however, that {\it all} pure states of ${\mathcal H}$ - 
whether they are conventionally unentangled or not relative to the selected 
state space decomposition - correspond to extremal reduced states relative 
to the full operator algebra $\mathfrak{g}= \mathfrak{su}(d)$: accordingly, 
all pure states of ${\mathcal H}$ are generalized unentangled relative to 
such algebra.  
Physically, this expresses the fact that no distinction between entangled 
and unentangled states is operationally meaningful if {\it full} access to 
arbitrary non-local operators is available.  In a sense, the emergence of 
entanglement in this generalized perspective appears intimately and directly 
associated with {\it physical constraints}.

\subsection{Multipartite qubit systems}

The relevant case of $n$ qubits is recovered by letting 
$d_\ell=d_0=2 \,\forall \ell$, in which case one simply has 
${\sf K}'=2/n$.    A natural choice for calculating 
the local purity via Eq. (\ref{ph}) is obtained by choosing standard
multi-qubit Pauli operators, 
\begin{equation} 
{\mathfrak{h}_{loc}} = \bigoplus\limits_{\ell =1}^n  \mathfrak{su}(2)_\ell = 
\text{span} \{ \sigma_x^1 , \sigma_y^1,
\sigma_z^1, \cdots , \sigma_x^n , \sigma_y^n , \sigma_z^n \},
\end{equation} 
where
\begin{equation}
\label{pauli3}
\sigma^\ell_\alpha = \overbrace{{\mathbb I} \otimes {\mathbb I} \otimes 
\cdots \otimes \underbrace{\sigma_\alpha}_{\ell^{th}\ \mbox{factor}} \otimes 
\cdots \otimes  {\mathbb I} }^{n\ \mbox{factors}} ,
\end{equation}
and the $2\times 2$ matrices $\sigma_\alpha$ and ${\mathbb I}$ are the 
standard Pauli matrices, satisfying $\sigma_\alpha^2={\mathbb I}$. 
By noticing that $x_\alpha= \sigma_\alpha/\sqrt{2}$ in this case, the 
purity of an arbitrary pure state becomes then
\begin{equation} 
P_{{\mathfrak{h}}} ( |\psi\rangle ) = \frac{1}{n} \sum\limits_{\alpha=x,y,z}
\sum\limits_{\ell =1}^n \langle \sigma_\alpha^\ell \rangle^2 .
\label{local}
\end{equation}
As above, $P_{{\mathfrak{h}}}=1$ in any product state of the form 
$|\psi\rangle = \otimes_\ell | {\psi_\ell}\rangle$, a multi-qubit GCS in 
this algebra.  Using Theorem 4.1, the relationships between local purities
and qubit reduced states rewrite respectively as follows:
\begin{equation}
P_{{\mathfrak{h}_\ell}} ( |\psi\rangle ) = 
2 \left(  {\rm{tr}} \rho_\ell^2 - \frac{1}{2} \right)
\label{hellreduced'}
\end{equation}
\begin{equation}
P_{{\mathfrak{h}_{loc}}} ( |\psi\rangle ) =  
\frac{2}{n}  \sum_{\ell=1}^n \left(
{\rm{tr}} \rho_\ell^2 - \frac{1}{2} \right)=
\frac{2}{n} \sum_{\ell=1}^n {\rm{tr}} \rho_\ell^2 -1 .
\label{hreduced'}
\end{equation}

\begin{example}
With the above definitions, all pure product states of $n$ qubits have 
maximal purity that is, they are generalized unentangled.  On the opposite 
limit, states of the form $|{\sf GHZ}_{n}\rangle = 
( |\uparrow \uparrow \ldots \uparrow\rangle
+ | \downarrow \downarrow \ldots \downarrow \rangle )/\sqrt{2}$ or of the 
form $[( |{\uparrow \downarrow}\rangle - 
|{\downarrow \uparrow}\rangle )/\sqrt{2}]^{\otimes n/2}$ 
(for $n$ even) will be maximally entangled according to this measure
($P_{{\mathfrak{h}}}=0$).
\end{example}

Because GE relative to the local algebra takes contributions, as quantified
by the measure $P_{{\mathfrak{h}}}$, from correlations involving {\it all}
subsystems simultaneously, one could expect such a measure to contain
information about overall entanglement properties of the system (as opposed,
for instance, to the concurrence measure which is intrinsically 
bipartite~\cite{W}).  A measure $Q$ of {\em global} entanglement for pure
states of $n$ qubits was recently proposed by Meyer and Wallach in~\cite{MW}, 
and shown to exhibit correct invariance properties under local unitary operations.  
According to the original construction, the first step to define such a 
measure is to consider a family of maps $l_j(b)$ given by the following 
action in the computational basis (and extended by linearity): 
\begin{equation}
l_j(b) |{b_1, \cdots ,b_n} \rangle= \delta_{bb_j} | {b_1 ,\cdots ,\hat{b}_j,
\cdots ,b_n} \rangle,
\label{lj}
\end{equation}
where $b$ and $b_j$ are either $|0 \rangle$ or $|1 \rangle$, 
and $\hat{b}_j$ denotes the absence of the $j$th qubit in the resulting 
$(n-1)$-qubit state.  Let any pure state in $({\mathbb C}^2)^{\otimes n-1}$
be represented in the computational basis, and let $|{u}\rangle =\sum u_i 
|{\phi_i}\rangle$ and $|{v}\rangle= \sum v_i |{\phi_i}\rangle$ be two 
such states. If a distance function on $ ({\mathbb C}^2)^{\otimes n-1} 
\otimes ({\mathbb C}^2)^{\otimes n-1}$ is introduced as 
\begin{equation}
D(u,v) = \frac{1}{2} \sum\limits_{i,j} |u_iv_j - u_jv_i|^2,
\label{dist}
\end{equation}
the global entanglement of a pure $n$-qubit state $|\psi\rangle$ is defined 
as follows:
\begin{equation}
\label{meyer}
Q( |{\psi}\rangle) = \frac{4}{n} \sum\limits_{j=1}^n D\Big(l_j( 0) |\psi\rangle,
l_j(1) |\psi\rangle \Big)\;,
\end{equation}
Interestingly, as explained in~\cite{BKOSV,SOBKV}, the measures $P_{{\mathfrak{h}}}$
and $Q$ turn out to have a simple relationship.

\begin{theorem} Let ${\mathfrak{h}}$ denote the local observable algebra for 
$n$ qubits as defined above.  Then for every pure state $|\psi\rangle \in 
{\mathcal H}= ({\mathbb C}^2)^{\otimes n}$ the following identity holds:
\begin{equation}
\label{pur}
P_{{\mathfrak{h}}} ( |\psi\rangle ) = 1-Q (|\psi\rangle )
\end{equation}
\end{theorem}
\begin{proof}
By partitioning the $n$ qubits into the $j$th qubit and the remaining one, and by 
working in the computational bases $\{|{0}_j \rangle , |{1}_j \rangle \}$ and 
$\{ |{\phi_i}\rangle \}$, $i=1,\ldots,2^{n-1}$ of the $j$th qubit and the rest,
respectively, any pure quantum state $|\psi\rangle \in 
({\mathbb C}^2)^{\otimes n}$ can be represented as
\begin{equation}
|{\psi}\rangle =\sum\limits_{i=1}^{2^{n-1}} \Big[g_i^j |{0}_j \rangle 
+ h_i^j |{1}_j \rangle \Big] |{\phi_i}\rangle , \hspace{5mm} 
j \in \{ 1,\ldots, n \}, 
\end{equation}
where $g_i^j$ and $h_i^j$ are complex coefficients. 
Therefore, the action of the map $l_j(b)$ on $|{\psi}\rangle$ given 
in Eq. (\ref{lj}) is 
\[
\begin{array}{l}
l_j(0) |\psi\rangle =  \sum\limits_{i=1}^{2^{n-1}} g_i^j 
|{\phi_i} \rangle  \\
l_j(1) |{\psi}\rangle = \sum\limits_{i=1}^{2^{n-1}} h_i^j 
|{\phi_i}\rangle. 
\end{array}
\]
Using Eq. (\ref{dist}), the distance between the above states of $(n-1)$ 
qubits becomes 
\begin{equation}
D\Big( l_j(0)|\psi\rangle ,l_j(1)|\psi\rangle \Big) 
=\sum\limits_{i,i'} \Big[ |g_i^j|^2
|h_{i'}^j|^2 -  (g_i^j h_{i'}^j) (h_i^j g_{i'}^j)^* \Big], 
\label{dd}
\end{equation}
where $^*$ denotes complex conjugate. After some algebraic manipulations 
we get
\begin{eqnarray*}
\sum\limits_{i=1}^{2^{n-1}} |g_i^j|^2 = \langle \psi | \left( \frac
{1+\sigma_z^j}{2} \right) |\psi \rangle, \\ 
\sum\limits_{i=1}^{2^{n-1}} |h_i^j|^2 = \langle \psi | \left( \frac
{1-\sigma_z^j}{2} \right) |\psi \rangle, \\
\sum\limits_{i=1}^{2^{n-1}} g_i^j (h_i^j)^*= \langle 
\psi | \sigma_-^j |\psi\rangle, \hspace*{9mm} 
\end{eqnarray*}
thus Eq. (\ref{dd}) rewrites as 
$$D\Big(l_j(0)|\psi\rangle ,l_j(1)|\psi\rangle \Big) =\frac{1}{4} 
\Big[1 - \langle \sigma_z^j \rangle^2 - \langle \sigma_x^j 
\rangle^2 - \langle \sigma_y^j \rangle^2\Big].$$ 
By taking a sum over all qubits and recalling Eq. (\ref{meyer}), we finally 
obtain the desired result
$$ Q( |{\psi}\rangle) = 1 - \frac{1}{n} \sum\limits_{j=1}^n \Big[ 
\langle \sigma_z^j \rangle^2  + \langle \sigma_x^j \rangle^2 + 
\langle \sigma_y^j \rangle^2 \Big] = 
1 - P_{ {\mathfrak{h}} } (|{\psi}\rangle). $$
\end{proof}

Theorem 4.5 together with Eq. (\ref{hreduced'}) also implies a direct, 
linear relationship between the global entanglement metric $Q$ of~\cite{MW} 
and the average subsystem purity, that is,
\begin{equation}
Q(|\psi\rangle) =
2 \Big( 1 - \frac{1}{n} \sum_{\ell=1}^n {\rm tr} \rho_\ell^2 \Big) .
\end{equation}
The above relationship has been independently established by Brennen~\cite{BR}.

\section{Generalized entanglement and subsystems}

According to the results of the previous section, GE includes conventional 
entanglement as a special case: when a given partition of ${\mathcal H}$ into 
subsystems is specified, there exists a natural observable subspace 
(the one describing local actions on individual factors) such that 
entanglement relative to the given subsystem partition and GE relative to 
the associated local algebra coincides for all states.  Suppose, however, 
that the opposite point of view is taken, namely that the state space 
${\mathcal H}$ of a quantum system is known, along with a distinguished, 
irreducible Lie algebra ${\mathfrak{h}}$ (or, more generally, an observable 
set $\Omega$).  
Then the following questions arise: Are there preferred subsystem 
partitions on ${\mathcal H}$ which may be determined by ${\mathfrak{h}}$?  
If so, does a (possibly fictitious) subsystem partition exist, such that 
generalized entangled states relative to ${\mathfrak{h}}$ are exactly the 
set (or are a subset) of ordinary entangled states relative to such a 
partition? In other words, does GE represent a genuine extension of the 
subsystem-based entanglement framework in its most general formulation? 

From the {\it physical} point of view, the novelty and added generality 
afforded by the GE notion should be already clear from the {\it direct} 
applicability of our prescription to any situation where a laboratory 
condition or physical constraint singles out a special observable set, 
shortcutting the need for any intermediate subsystem identification.  
From a {\it mathematical} perspective, a distinctive innovation with 
respect to the traditional framework is that GE ultimately depends only 
on the {\it convexity} properties of observable sets~\cite{BKOV}.  The
the resulting extremality characterization of unentangled states has not 
been emphasized in the conventional setting (for instance, it is not 
part of ~\cite{ZLL}'s treatment of the virtual-subsystem approach).  
With these general considerations in mind, however, a thorough analysis 
of the relationships between our approach and the subsystem-based one is 
an interesting issue worth investigating {\it per se}.  
We limit ourselves here to a few illustrative remarks.  

\subsection{Observable-dependent subsystems}

If, as assumed so far, the distinguishable observables form a semisimple
Lie algebra ${\mathfrak{h}}$, a natural multipartite setting is determined 
by the fact that ${\mathfrak{h}}$ is uniquely expressible as a product of
simple Lie algebras ${\mathfrak{h}}= \times_k {\mathfrak{h}}_k$~\cite{HU}. 
A Hilbert space irreducibly representing ${\mathfrak{h}}$ then factors 
as\footnote{Note that this implies $d$=dim(${\mathcal H}$) 
to be a non-prime integer.}
${\mathcal H}=\otimes_k {\mathcal H}_k$, 
with the algebra ${\mathfrak{h}}_k$ acting as the identity over all 
factors except the $k$th one.  In the language of~\cite{ZLL}, the 
resulting multipartition is formally reminiscent of an {\it 
observable-induced} tensor product structure associated to 
$\{ {\mathfrak{h}}_k \}$ over ${\mathcal H}$. Similarly, 
a formal analogy with the emergence of encoded multipartitions 
(${\mathcal C} \subset {\mathcal H}$) may be expected for a reducible
representation of $ {\mathfrak{h}}$ on ${\mathcal H}$.  Note, however, 
that actions on the individual state spaces ${\mathcal H}_k$ belong to
a Lie group representation which need {\it not} be 
GL(dim(${\mathcal H}_k$)) as for standard entanglement theory~\cite{BKOV}.

\subsection{Entanglement without subsystems}

A situation which strikingly illustrates the added flexibility of 
the GE notion is offered by quantum systems whose states space is
intrinsically irreducible, that is, $d$ is a prime number.  In this 
case, the system is physically elementary, and conventional 
entanglement is not directly applicable. Consider for instance a 
single spin-1 system, whose three-dimensional state space 
${\mathcal H}= {\mathbb C}^3$ carries an irreducible representation 
of $su(2)$, with usual angular momentum generators $J_x, J_y, J_z$ 
given by
\begin{equation}
J_x=\frac{1}{\sqrt{2}} \begin{pmatrix}
0 & 1 & 0 \cr
1 & 0 & 1 \cr
0 & 1 & 0 
\end{pmatrix}, \; 
J_y=\frac{1}{\sqrt{2}} \begin{pmatrix}
0 & -i & 0\cr 
i & 0 &-i \cr
0 & i& 0 ,
\end{pmatrix}, \;
J_z=\frac{1}{\sqrt{2}} \begin{pmatrix}
1 & 0 & 0\cr
0 & 0 & 0\cr
0 & 0 & -1 
\end{pmatrix}.
\end{equation}
Suppose that, due to operational limitations, observations on this 
system are restricted to observables linear in the above generators
of ${\mathfrak{su}}(2)$. Given an arbitrary pure state $|\psi\rangle 
\in {\mathbb C}^3$, the corresponding reduced state can be identified 
with the vector of expectation values of these three observables 
$\langle J_\alpha \rangle$.  The set of such reduced states is a unit 
ball in ${\mathbb R}^3$, defined by $\langle J_x\rangle^2+\langle
J_y\rangle^2+\langle J_z\rangle^2\leq 1$.  The extremal states correspond
to points on the surface, resulting from maximal spin component $1$ for 
some linear combination of $J_x, J_y, J_z$.   For any choice of spin
direction, ${\mathcal H}$ is spanned by the $|1, +1\rangle, |1, 0\rangle,
|1, {-1}\rangle$ eigenstates of that spin component, for instance the 
$\hat{z}$ direction.  Then the extremal states are immediately 
identified with the {\it spin coherent states} $|1,\xi\rangle$, or 
GCSs for SU(2)~\cite{ZFG},
\begin{equation}
|1, \xi\rangle = \text{e}^{\xi J_+ - \xi^\ast J_-} |1,-1\rangle,
\hspace{5mm} \xi \in {\mathbb C},   
\end{equation}
where the exponential involving the ladder operators $J_\pm$ provides 
an explicit realization of the group-displacement ${\mathcal D}$ of Eq. 
(\ref{gcs}), and the lowest-weight state $|J=1, J_z=-1\rangle$ is 
chosen as the reference state. Note that the states $|1, +1\rangle$ and 
$|1, -1\rangle$ are GCSs, but $|1, 0\rangle$ is {\it not}: thus, the 
latter is a generalized (in fact, maximally) entangled state relative 
to ${\mathfrak{su}(2)}$.  
As also remarked earlier, access to the {\it full} operator algebra 
(${\mathfrak{su}(3)}$ in this case) causes GE to disappear altogether, 
expressing the fact that arbitrary state vectors can be connected 
through group transformations in SU(3). 

Note that, in principle, one could formally regard the single 
spin-$1$ system considered above as arising from two spin-$1/2$ 
subsystems in a triplet configuration.  Then states which are 
generalized unentangled (or entangled) relative to ${\mathfrak{su}(2)}$
would be associated with ordinary product (or entangled) states of the
two fictitious subsystems, 
\begin{eqnarray*}
|1, {-1}\rangle & \leftrightarrow &  
|\downarrow\rangle_1 \otimes |\downarrow\rangle_2 \\
|1, {0}\rangle & \leftrightarrow &  
\frac{ |\uparrow \rangle_1 \otimes |\downarrow\rangle_2 + 
|\downarrow \rangle_1 \otimes |\uparrow\rangle_2 }{\sqrt{2}} \\
|1, {+1}\rangle & \leftrightarrow &  
|\uparrow\rangle_1 \otimes |\uparrow\rangle_2,
\end{eqnarray*}
with self-explaining notation.
We would like to stress that, from a physical point of view, such a partition 
is entirely artificial if operational access to the individual subsystems is 
unavailable as assumed above.  Moreover, the procedure is everything but 
straightforward mathematically.  Formally, the above line of reasoning 
requires embedding the original spin-$1$ irreducible representation of 
${\mathfrak{su}(2)}$ in the tensor product representation of two spin-$1/2$ 
irreducible representations of ${\mathfrak{su}(2)}$, in such a way that 
generalized entangled states becomes a {\it subset} of the ordinary 
entangled states in the extended representation\footnote{Clearly, the 
(conventionally) entangled state spanning the singlet sector has {\em no} 
counterpart in the original physical space ${\mathcal H}$.}.  While the 
simplicity of the state identification given above is coincidental to the 
elementary irreducible-representation-structure of this example, understanding 
to what extent a similar procedure could be carried out for general irreducible 
representations of semisimple Lie algebras might shed further light on the 
mathematical relationships between GE and abstract subsystem structures
(See also discussion in Section 5).

\section{Discussion}

\subsection{Implications for condensed-matter systems}

A feature which makes the GE framework particularly attractive for
applications to condensed-matter systems is the possibility to 
directly formulate entanglement in terms of the algebraic operator 
language (fermionic, bosonic, or other~\cite{BO}) which best 
describes the system.  Among such applications, the possibility
of gaining a better understanding of the nature and properties
of the quantum correlations in a system undergoing a 
{\it quantum phase transition} has attracted a growing interest
recently, see for instance~\cite{ON,OAFF,VLRK}.  The usefulness 
of the GE framework in the context of identifying and characterizing
quantum phase transitions has been investigated based on the explicit 
analysis of two integrable models undergoing a broken symmetry quantum 
phase transition -- the well known one-dimensional spin-1/2 XY model in 
a transverse magnetic field and the so-called Lipkin-Meshkov-Glick model,
respectively.  We refer the interested reader to ~\cite{BKOSV,OSBKV,SOBKV} 
for detailed discussions.  As these studies reveal, the purity relative 
to an appropriate Lie subalgebra of observables of the system provides a 
useful diagnostic tool for characterizing the many-body correlations
which play a dominant role at criticality, by succeeding at both 
sharply detecting the critical point and correctly identifying the 
underlying universality class.  

While the above results are very suggestive and promising, several 
important questions remain to be investigated in more depth.  In 
particular, a general operational criterion for identifying the 
physically relevant observable subalgebra for Hamiltonians whose 
ground-state properties are not easily computable is still lacking. 
The identification of a {\it minimal} collection of observable algebras 
able to provide a complete description of the system's critical properties 
is likewise an open problem.  Finally, the extension of similar concepts 
and techniques to more general classes of quantum phase transitions 
(notably, topological quantum phase transitions) seems likely to require
more sophisticated tools than the Lie-algebraic ones which suffice
when a broken symmetry exists.

\subsection{Outlook and conclusion}

We have presented a generalization of entanglement which provides a 
{\it subsystem-independent}, unifying conceptual framework for defining 
entanglement in arbitrary physical settings.  Unlike the conventional
definition which is relative to a preferred decomposition into 
subsystem, GE is directly regarded as an {\it observer-dependent} 
property of quantum states, which is definable relative to any 
physically relevant set of observables for the system. Whenever the 
latter possess a Lie-algebraic structure, our approach naturally links 
entanglement theory with the theory of generalized coherent states. 

In addition to the condensed-matter implications mentioned above, 
numerous information-theoretic problems also deserve further investigation.
Several issues concerning appropriate generalizations of local maps, 
resource scaling, and GE measures have been raised and partially 
addressed in~\cite{BKOV}.  Additional research directions might 
involve exploring possible connections between the presence of GE 
and the violation of Bell-type inequalities~\cite{PS}, assessing the 
potential of GE detection via appropriate witness operators~\cite{HO,T}, 
looking at possible characterizations of GE via uncertainty relations 
as recently suggested for conventional entanglement~\cite{HT}, and more.  
Ultimately, the fresh perspective offered by our approach 
will deepen our understanding of entanglement as a physical and 
information-theoretic resource.

\bibliographystyle{amsalpha}

\end{document}